# CONCEPTUAL DESIGN OF THE MUON COLLIDER RING LATTICE*

Y. Alexahin, E. Gianfelice-Wendt, A. Netepenko, FNAL, Batavia, IL 60510, U.S.A.

*Abstract*

Muon collider is a promising candidate for the next energy frontier machine. However, in order to obtain peak luminosity in the $10^{35}/\text{cm}^2/\text{s}$ range the collider lattice design must satisfy a number of stringent requirements, such as low beta at IP ($\beta^* < 1$ cm), large momentum acceptance and dynamic aperture and small value of the momentum compaction factor. Here we present a particular solution for the interaction region optics whose distinctive feature is a three-sextupole local chromatic correction scheme. Together with a new flexible momentum compaction arc cell design this scheme allows to satisfy all the above-mentioned requirements and is relatively insensitive to the beam-beam effect.

## INTRODUCTION

Large values of the transverse emittance and momentum spread in muon beams that can be realistically obtained with ionization cooling impose unprecedented requirements on the muon collider optics, the major problem being the interaction region (IR) quadrupoles chromaticity correction. The lattice design is also complicated by the necessity to protect magnets and detectors from products of muon decay, these issues are addressed in the accompanying paper [1].

In our earlier report [2] we reviewed previous attempts to design a muon collider optics and proposed a new approach to chromaticity correction which apparently succeeded in achieving the main goals: large momentum acceptance and dynamic aperture.

In the present report we describe further development of that approach, complement the IR design with new arc optics and study beam dynamics with account of the beam-beam effect and magnet nonlinearities.

## LATTICE DESIGN

The muon beam parameters assumed for the lattice design are given in Table 1. We impose rather conservative limits on magnet strength: B=8 T for dipoles at high-beta locations and 10 T in the arcs, 250 T/m for quadrupoles with 80 mm aperture and proportionally lower for larger apertures (see [1] for detail). The distance from IP to the first quad is set at 6 m.

### IR Chromaticity Correction

The IR design proposed in [2] borrowed heavily from the 1996 design by K. Oide [3]. According to the latter the vertical $\beta$-function after the final focus (FF) doublet is much larger than the horizontal one (~900 km for $\beta^* = 3$ mm). Accordingly larger vertical chromatic

_________
* Work supported by Fermi Research Alliance, LLC under Contract DE-AC02-07CH11359 with the U.S. DOE.
# alexahin@fnal.gov

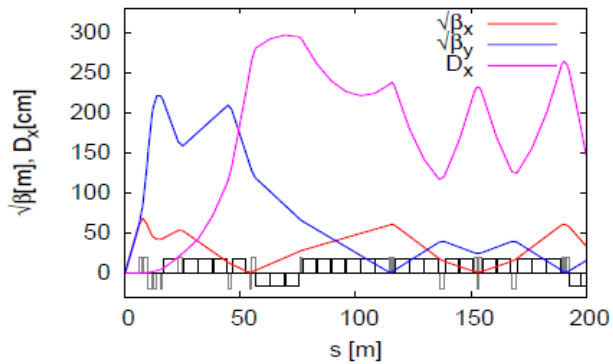

Figure 1 (color): IR layout and optics functions function is corrected first with a pair of sextupoles separated by a –I section. The horizontal chromatic function is corrected with another pair of sextupoles, so there is a total of four sextupoles on each side of the IP.

The scheme proposed in [2] – whose new version is shown in Fig. 1 – differs from [3] in that there is just one sextupole for vertical chromaticity correction on each side of the IP whose neighbourhood serves as a –I section. Thus the number of key sextupoles on each side of the IP is reduced to three.

Since the vertically correcting sextupoles are at the same vertical betatron phase as the generating chromatic $\beta$-wave FF quads, the scheme is quite robust with respect to field errors. However, the beam-beam effect has to be compensated right at the IP for such a scheme to work with colliding beams. This would require implementation of as-yet-untested methods like plasma beam-beam compensation.

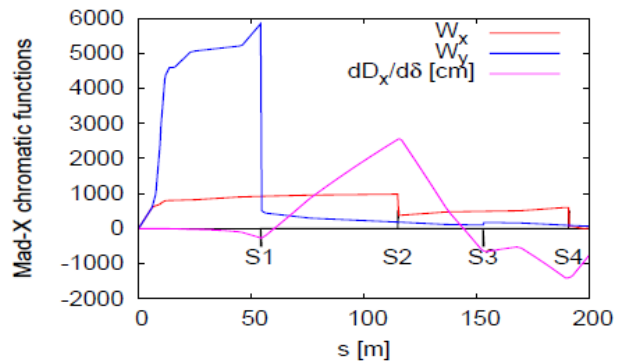

Figure 2 (color): IR chromatic functions and sextupoles location.

It turned out that sensitivity to the beam-beam effect and the overall performance can be significantly improved by dropping the condition of 180° horizontal phase advance between the vertically correcting sextupoles and placing them at the minima of the horizontal $\beta$-function (actually this required moving them by just a few meters). Since the resonance driving terms and detuning coefficients due to normal sextupoles all include powers of $\beta_x$, they can be made sufficiently small in this way without resorting to error-prone $-I$ sections.

Smallness of $\beta_y$ at location of the horizontal chromaticity correction sextupoles is beneficial but not sufficient for suppression of aberrations they produce, so a $-I$ separated pair is still necessary (marked as S2 and S4 in Fig. 2). The dispersion function at the center of the corresponding $-I$ section can be made rather large while $\beta_x$ is small (~1 m in the present design with $\beta^*$ =1 cm) making it a convenient place for one more sextupole (S3) which provides control of the second order dispersion. Multipoles for higher order vertical chromaticity correction can be installed there as well.

The described scheme is rather insensitive to the beam-beam effect since the $\beta$-wave excited by the beam-beam interaction reduces $\beta_x$ at S1 and S3 locations and $\beta_y$ at S2 and S4 locations suppressing the aberrations even more.

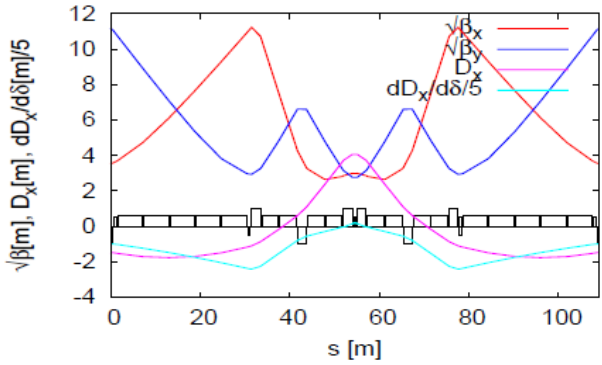

Figure 3 (color): Layout and optics of an arc cell.

### Arc Cell

The interaction region produces large positive contribution to momentum compaction factor $\alpha_p$ which must be compensated by a negative contribution from the arcs.

There is a number of arc cell designs which provide negative $\alpha_p$. The KEKB design used in [3] has some drawbacks: the ratio of $\beta$-functions at locations of sextupoles for horizontal and vertical chromaticity correction is rather small so that they fight each other and have to be too strong; there is no independent "knob" for correction of $\alpha_p$ derivative with relative momentum deviation $\delta_p$.

In the ultra-relativistic case this derivative is given by expression

$$\frac{d\alpha_p}{d\delta_p} = \frac{1}{C}\int_0^C\left[\frac{1}{\rho}\frac{dD_x}{d\delta_p} + \frac{1}{2}\left(\frac{dD_x}{ds}\right)^2\right]ds \quad (1)$$

where $\rho$ is the radius of orbit curvature and $C$ is the machine circumference. Since the second term in the integrand is always positive a negative second-order dispersion is necessary to cancel it.

Figure 3 shows the arc cell design which permits an independent control of all important parameters. Since both $\beta$-functions are small at the cell center while $D_x$ is at its maximum, the quadrupoles and a sextupole located there control $\alpha_p$ and $d\alpha_p/d\delta_p$ respectively without big impact on phase advances and chromaticity.

The arc sextupoles are relatively weak so there is no need to organize them in non-interleaved pairs. We chose the betatron phase advance 300° per cell in both planes to get the 3rd order resonances cancelled in each arc consisting of 6 cells.

With 2 IRs and short dispersion-free matching sections containing RF cavities the full circumference is 2.5 km.

### Tuning section

The design should be flexible enough to accept a wide range of muon beam emittances since there is an uncertainty in the value which can be obtained in the ionization cooling channel. With higher emittance $\beta$-functions in the IR magnets should be smaller in order to accommodate the beam inside limited aperture, in the result $\beta^*$ has to be larger. Conversely, with lower emittance smaller values of $\beta^*$ are allowed.

A special section was designed which allows $\beta^*$ variation in the range 0.5-2 cm as well as the betatron tune adjustment without any changes in IR and the arc. However, this section came out rather long increasing the circumference to 2.727 km.

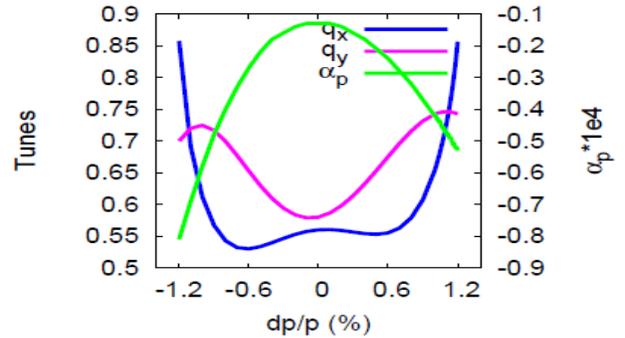

Figure 4 (color): Fractional parts of the tunes and momentum compaction factor vs. momentum deviation.

## LATTICE PERFORMANCE

To correct nonlinear chromaticity and detuning with amplitude four octupoles and two decapoles were added on each side of the IP. The resulting dependence of the betatron tunes and momentum compaction factor on momentum is shown in Fig. 4, the central values being $Q_x$=20.56, $Q_y$=16.58, $\alpha_p$=-1.3·10$^{-5}$. The static momentum acceptance of ±1.2% by far exceeds the requirements of the baseline muon collider scheme.

The dynamic aperture (DA) was computed with MAD8 by tracking particles for 1000 turns. Figure 5 shows initial positions of stable and lost particles in the plane of Courant-Snyder amplitudes for $\beta^*$=1 cm. Conventional DA can be calculated as $A_{min}\cdot(\gamma/\varepsilon_{\perp N})^{1/2}$ and amounts to 5.7σ for nominal emittance cited in Table 1. This exceeds the design aperture of 5σ in the FF quads.

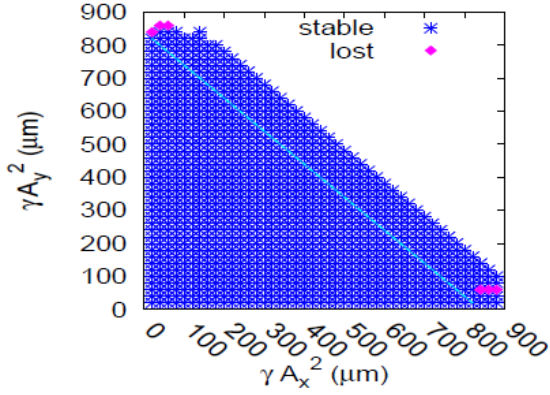

Figure 5 (color): 1000 turns dynamic aperture

With $\beta^*$ variation in the range 0.5-2 cm (and proportional variation of the normalizing emittance) the dynamic aperture measured in sigmas stays approximately constant.

Synchrotron oscillations reduce the DA: it reaches the marginal value of $2.5\sigma$ at synchrotron amplitude $\delta_p > 0.5\%$ which means that the beam r.m.s. momentum spread is limited by 0.2% which is twice larger than the design value.

In these studies the effect of fringe fields was not taken into account though it may be quite important [3]. Also important are nonlinear components in magnets, particularly strong in IR dipoles [4]. Preliminary analysis shows that the effect of these components can be compensated with dedicated correctors.

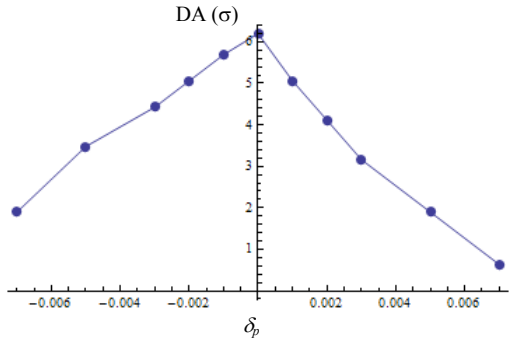

Figure 6: Dynamic aperture vs. momentum deviation in the presence of beam-beam effect ($\xi$ = 0.1/IP)

*Beam-Beam Effect*

We chose the working point above half-integer to minimize detuning with amplitude and orbit sensitivity to misalignments and to provide space for the beam-beam tuneshift. As for the beam-beam effect itself, it is considered beneficial to have phase advance between IPs above a multiple of $\pi$ which would make an integer tune in our case of two IPs. But strong "dynamic beta" effect increases $\beta$-functions at the FF quads and therefore should be avoided.

Still, with the chosen tunes there is some reduction in $\beta^*$ for long bunches, $\sigma_s \approx \beta^*$, which almost completely compensates for the "hour-glass" effect at beam-beam parameter values $\xi \approx 0.1/IP$. Interestingly enough, $\beta$-functions at the FF quads are also reduced in this case.

Detailed studies of beam dynamics with account of the beam-beam effect will be reported elsewhere. Figure 6 shows quite encouraging preliminary results for 1000 turns "diagonal" dynamic aperture (along the line $A_x=A_y$) at constant momentum deviation.

## SUMMARY

The developed in this report "three-sextupole" scheme for IR chromaticity local correction and new arc cell configuration allowed us to achieve the goal parameters of muon collider design summarized in Table 1.

Table 1: Muon Collider Parameters

| Beam energy | TeV | 0.75 |
|---|---|---|
| Number of IPs | - | 2 |
| Circumference, $C$ | km | 2.73 |
| $\beta^*$ | cm | 0.5-2 |
| Momentum compaction, $\alpha_p$ | $10^{-5}$ | -1.3 |
| Normalized emittance, $\varepsilon_{\perp N}$ | $\pi$·mm·mrad | 25 |
| Momentum spread | % | 0.1 |
| Bunch length, $\sigma_s$ | cm | 1 |
| Number of muons / bunch | $10^{12}$ | 2 |
| Beam-beam parameter / IP | - | 0.09 |
| Dynamic aperture | $\sigma$ | 5.7 |
| Static momentum acceptance | % | ±1.2 |
| Average luminosity / IP | $10^{34}/cm^2/s$ | 1.1 |

## ACKNOWLEDGEMENTS

The authors are grateful to R. Palmer and A. Tollestrup for many useful remarks and to Dr. K. Oide for kindly providing the detail of his MC design.